\documentstyle[prl,epsf,aps]{revtex}

 \def\(({\left(}
 \def\)){\right)}

\def \beq{\begin{equation}}
\def \eeq{\end{equation}}

\def \ab2{\alpha\beta^2}

\newcommand {\be} {\begin{equation}}
\newcommand {\bea} {\begin{eqnarray} \nonumber }
\newcommand {\ee} {\end{equation}}
\newcommand {\eea} {\end{eqnarray}}
 \newcommand {\eps} {\epsilon}

\newcommand {\lan} {\langle}
\newcommand {\ran} {\rangle}

\begin{document}
\input{epsf}
\twocolumn \vskip.5pc
\narrowtext 
{\bf Comment on ``Dynamical Heterogeneities in a Supercooled Lennard-Jones
Liquid''}
\vspace{0.5 cm}

In two recent interesting 
letters \cite{LORO} 
evidence was presented for the existence of a growing  dynamic 
correlation length when we approach the glass transition from the 
liquid phase (a similar 
divergence is present also in the off-equilibrium dynamics below $T_c$ 
\cite{PARI}).  Here we would 
like to point out that this phenomenon can be easily predicted
 using the replica approach of 
reference \cite{FP}.   Let us call $\rho(x,t)$ the density at time $t$ 
($\rho(x,t)=\sum_i
\delta(x_i(t)-x)$).  It is convenient to consider the quantity
$ Q_{z_1}(x,t)
 = \rho(x-z_1/2,0) \rho(x+z_1/2,t)$.
Using it we can construct the  correlation
$C(x-y,t)=
\lan Q_{z_1}(x,t)Q_{z_2}(y,t) \ran $
that should be weakly depend on the values of 
$z_{1}$ and $z_{2}$ for  $|z_1|,|z_2|<<x-y$.  The data in \cite{LORO} 
can be read as evidence for the existence of a correlation length 
that controls the decay of $C$ at large spatial argument. 
This length reaches a maximum at 
a time $t$ comparable with the 
relaxation time of the system and this time grows for lower temperatures.  Let us take the mode 
coupling point of view, assuming an incipient divergence of the relaxation time and breaking of 
ergodicity at a temperature $T_c$.  Exactly at the mode coupling transition we would sample at large 
times a single ergodic component and obtain
\bea
& C_\infty(x)=\lim_{t \to \infty} C(x-y,t)= \\
&\lan \rho(x+z_1/2)\rho((y+z_2/2)\ran_{c} \lan \rho(x-z_1/2)\rho(y-z_2/2\ran _c
\eea
where the expectation value is taken inside the same ergodic component and $c$
 stands for connected.  
As soon as we approach $T_c$ and the times become large, the dynamic 
correlation function goes, in 
the appropriate time window, to the critical one. Our burden is to
prove that there are (within the approximation used) long range 
correlations at the mode coupling point.   
\begin{figure}
\epsfxsize=200pt \epsffile{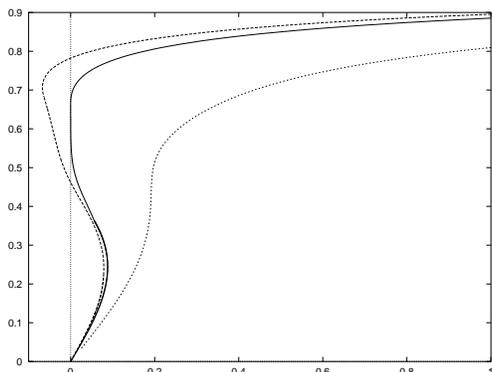}   
\caption[0]{\protect\label{F_E3} We plot $\partial W/\partial \eps$
as a function of the coupling $\eps$ for the model in \cite{FP} and 
$T=T_c(1-0.1)$ (dashed line), $T=T_c$ (full line) 
and $T=T_c(1+0.5)$ (dotted line). Notice the divergence of the $\partial^2
W/\partial \eps^2$ for $\eps\to 0$ for $T=T_c$ in the high 
branch of the curve.} 
\end{figure} 
In \cite{FP} it was shown that the mode coupling transition may be 
characterized as follows.  
Consider a system that thermalizes at temperature $T$ with 
an attractive coupling $\eps$ 
with a reference configuration which is itself 
a fixed equilibrium configuration at temperature $T$. We call $W$
the free-energy as a function of $\eps$. 
 There are three possibility for the expectation value of 
$(\partial W/\partial \eps)|_{\eps=0}$: (a) 
A non zero stable value below the ideal glass transition ($T<T_K$)
(b) Flatly a zero value above the Mode Coupling transition ($T>T_c$)
(c) A metastable non zero value in the 
region $T_K<T<T_c$.  The metastable state exists also for 
non zero coupling and is destabilized on the
spinodal line.  At $T_c$ the spinodal point  is at $\eps=0$.  
The situation is  sketched in fig 1.
It is a general feature of spinodal points
 that  the second derivative 
of the free energy, and its associated correlation length are divergent.
This is precisely the length associated to $C_\infty(x)$.
The learned reader may notice that we are just finding 
what one should expect from 
 the ``principle of marginal stability'' \cite{horner} 
which states
that the mode coupling transition point is critical.
Similar conclusions follows also from the 
geometrical description of \cite{KuLa}. 
It should also be evident that in principle this  correlation 
length  could be computed directly in the mode coupling 
approach. 
Unfortunately such a computation is 
not so simple and the method of \cite{FP} provides an useful short cut.
We stress that the divergence is present only in  the average 
of the product of two 
correlation functions inside the same ergodic component. 
No divergent length would  be seen in equilibrium 
correlation functions, whose dominant contribution comes from 
configurations in different ergodic components.

Silvio Franz(*) and Giorgio Parisi(**)\\
(*) Abdus Salam  ICTP
Strada Costiera 11,
P.O. Box 563,
34100 Trieste (Italy)\\
(**) Universit\`a di Roma ``La Sapienza''
Piazzale A. Moro 2, 00185 Rome (Italy)

\end{document}